# Cyberattack Data Analysis in IoT Environments using Big Data

Neelam Patidar, Sally Zreiqat, Sirisha Mahesh, Jongwook Woo
Department of Computer Information Systems,
California State University, Los Angeles
{npatida, szreiqa, smahesh4, jwoo5}@caltstatela.edu

**Abstract:** In the landscape of the Internet of Things (IoT), transforming various industries, our research addresses the growing connectivity and security challenges, including interoperability and standardized protocols. Despite the anticipated exponential growth in IoT connections, network security remains a major concern due to inadequate datasets that fail to fully encompass potential cyberattacks in realistic IoT environments. Using Apache Hadoop and Hive, our in-depth analysis of security vulnerabilities identified intricate patterns and threats, such as attack behavior, network traffic anomalies, TCP flag usage, and targeted attacks, underscoring the critical need for robust data platforms to enhance IoT security.

## 1. Introduction

The Internet of Things (IoT) is reshaping industries like healthcare, transportation, and industrial applications through interconnected networks of sensors and actuators. However, these advancements pose significant challenges, particularly in security, interoperability, and standardization. The pressing nature of these challenges is underscored by the fact that existing IoT attack datasets often lack comprehensive coverage and realistic environments, posing a significant hurdle to developing effective security solutions. Our research aims to address these urgent security challenges and contribute to creating a more secure IoT environment.

Our research employs the CICIoT2023 dataset to address these gaps and foster the development of robust IoT security analytics. We analyzed this data because of the critical need to secure the rapidly growing numberof IoT devices increasingly targeted by sophisticated cyberattacks. The CICIoT2023 dataset includes data from 33 attacks categorized into seven types (DDoS, DoS, Recon, Web-based, brute force, spoofing, and Mirai) across a network topology of 105 IoT devices focusing uniquely on attacks executed by malicious IoT devices and reflecting on real-world scenarios [2].

We aim to transform complex datasets into actionable insights using Apache Hadoop and Hive for data processing and Microsoft Power BI, Tableau, and Excel for visualization. This work is important as it advances secure IoT operations and contributes to developing effective security solutions, ultimately fostering a safer IoT ecosystem.

## 2. Related Work

Several studies have developed IoT security datasets and methodologies that parallel our research. The N-BaIoT dataset by Meidan et al. focuses on detecting botnet attacks using Mirai and BASHLITE botnets, with deep-learning autoencoders analyzing network traffic features. The BoT-IoT dataset by Koroniotis et al. captures realistic traffic, including DDoS and data theft attacks, and is evaluated with various machine and deep learning models. The IoT-23 dataset by Firdaus et al. emphasizes botnet attack data in real network environments for developing intrusion detection systems [3].

Despite these contributions, our work differs significantly in several key aspects:

- Big Data Framework: Our research uses Apache Hadoop and Hive on Cloud Computing platforms, enabling scalable and flexible processing of large-scale datasets. This is crucial for handling vast data volumes typical in real-world IoT environments.
- Comprehensive Dataset: The CICIoT2023 dataset provides a more extensive coverage of potential IoT attacks.
- Focus on IoT Devices as Attackers: Unlike many studies, our dataset includes attacks by malicious IoT devices, highlighting specific vulnerabilities in the IoT ecosystem.
- Advanced Data Visualization: We transform complex datasets into actionable insights using multiple visualization platforms; Microsoft Power BI, Tableau and Excel, aiding in pattern identification and security solution development.
- Realistic IoT Environment: The dataset is collected from a network topology of 105 real IoT devices, ensuring applicability to practical scenarios.
- Scalability and Flexibility: Our methodology allows for scalable analysis, accommodating larger datasets for deeper insights into IoT security.

By integrating these innovative approaches, our paper should contribute to advances in IoT security analytics, supporting a safer IoT ecosystem and setting it apart from existing studies.

## 3. Specifications

The CICIoT2023 dataset is designed to support the development of security analytics applications for real IoT operations. It includes detailed information on attacks executed by malicious IoT devices targeting other IoT devices. The data was collected from an extensive IoT topology comprising real IoT devices. Each attack scenario includes data captured from the perspective of both the attacker and the target devices. The original dataset is significant, totaling approximately 12.5 GB, which we reduced to 2 GB. The data is organized into several CSV files containing specific information about attacks and network traffic.

Table 1 shows the files and size of the files from the dataset.

*Table 1 Data Specification*

| | |
|---|---|
| Data Size | 2 GB |
| Number of Files | 30 |
| Data format | CSV |

Table 2 below shows our Oracle cluster and our Hadoop specifications.

*Table 2 H/W Specification*

| | |
|---|---|
| Number of nodes | 5 (2 master nodes, 3 worker nodes) |
| Cluster version | 3.3.3 |
| CPU Speed | 1995.312 MHz |
| Memory | 806.4 GB |
| Storage | 678 GB |

Table 3 below shows some dataset columns imported and used in our analysis.

*Table 3 Sample Dataset Columns*

| Column | Explanation |
|---|---|
| flow_duration | Time between first and last packet received in flow |
| Header_Length | Length of packet header in bits |
| Protocol Type | Protocol numbers as defined by IANA |
| Duration | Time-to-Live (ttl) |
| Rate | Rate of packet transmission in a flow |
| Srate | Rate of outbound (sent) packets transmission in a flow |
| Fin_flag_number | Fin flag value |
| Syn_flag_number | Syn flag value |
| Ack_flag_number | Ack flag value |
| HTTP | Indicates if the application layer protocol is HTTP |
| HTTPs | Indicates if the application layer protocol is HTTPS |
| DNS | Indicates if the application layer protocol is DNS |
| TCP | Indicates if the transport layer protocol is TCP |
| UDP | Indicates if the transport layer protocol is UDP |
| ICMP | Indicates if the network layer protocol is ICMP |
| label | Notes the type of attack being run or 'BenignTraffic' for no attack run |

## 4. Implementation Flowchart

Our analysis commenced with acquiring the CICIoT2023 dataset, which contains detailed information on various IoT attack types sourced from a reliable cybersecurity data repository. The implementation steps are outlined in the accompanying flowchart (Figure 1).

Initially, 30 data logs, formatted in CSV, were uploaded to the HDFS. After uploading, HiveQL was utilized to create the table schema, clean the data, generate summary tables, and export the results. After downloading the output files in CSV format, we utilized 3D column charts and pie charts in Power BI, Excel, and Tableau to create visualizations and generate insights.

*Figure 1 - Implementation Flowchart*

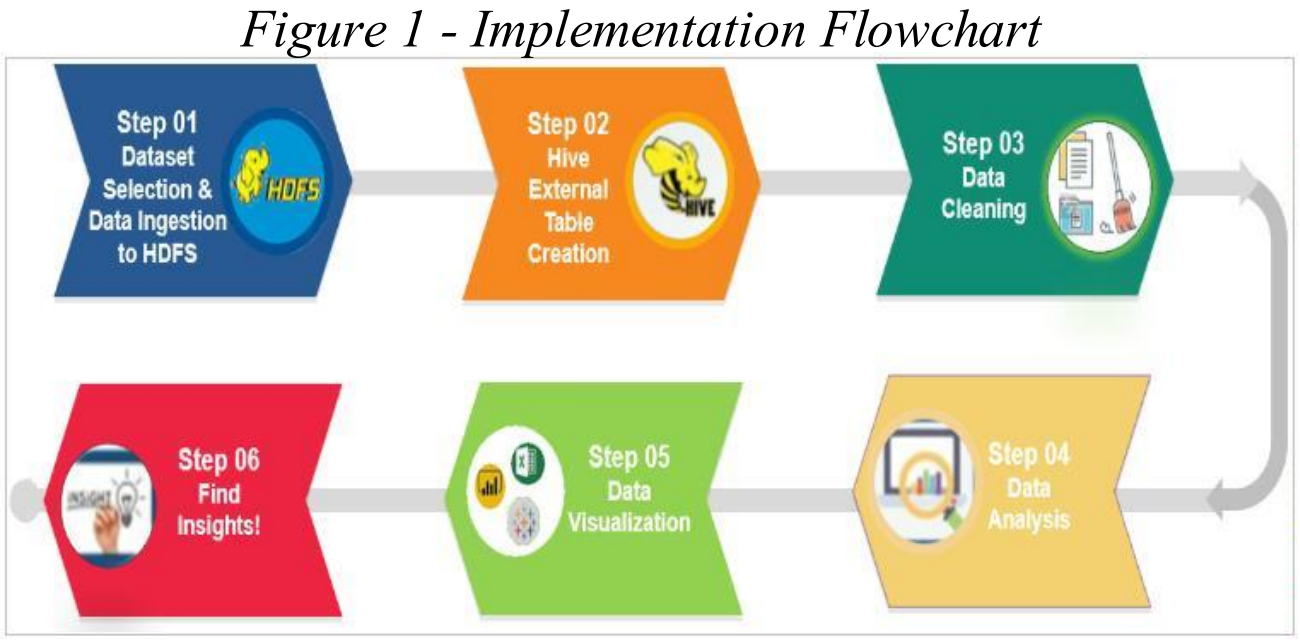

## 5. Data Cleaning

Data cleaning in our research involved several systematic steps to ensure the dataset was adequately prepared for analysis. Initially, we downloaded raw data files from their source website. Due to the extensive size and resource consumption, we merged and reduced their size before uploading them to the Hadoop Distributed File System (HDFS). We loaded the data into tables within the Hadoop ecosystem using the Beeline command-line interface.

'The dataset comprised multiple CSV files and tables with numerous columns. To streamline the analysis, we created a "clean" table containing only the specific columns necessary for our analysis, namely “IoT_Data,” achieved through Hive QL queries. Additionally, rows containing NULL values were removed to ensure data quality and usability.

During the data cleaning process, we addressed several challenges. The dataset included columns with binary data (0s and 1s) for protocols and flags, which required mapping to fetch the correct information for our queries. Since much of the table data was numeric and difficult to interpret, we identified columns that could establish relationships for all queries in the analysis. In addition, we have extracted categories from attack types using the HiveQL CASE statements.

## 6. Analysis and Visualization

Following the data-cleaning process, we conducted an in-depth analysis of the refined dataset. The cleaned dataset enabled us to examine the attacks in IoT environments, focusing on traffic metrics, protocol impacts, flag analysis in network traffic, and other critical aspects.

We employed advanced data visualization tools to gain insights into attack patterns and behaviors. Our goal was to present the data clearly and informatively. For this purpose, we utilized Microsoft Power BI, Tableau, and Excel, each offering unique capabilities contributing to a comprehensive understanding of the dataset.

Through these tools, we created visual representations highlighting key trends and correlations, facilitating a deeper understanding of the security challenges faced by IoT environments.

### 6.1 Distribution of Targeted Attacks by Category and Type

The first visualization, “Figure 2 - Attacks % per Category and Type,” was created using Microsoft Power BI and presented as a pie chart. This chart illustrates the distribution of targeted attacks across different categories and types within the IoT ecosystem. The parameters utilized include attack types, categories, and attack counts, enabling the identification of the most and least frequent attack occurrences. The analysis revealed that Distributed Denial of Service (DDoS) attacks were the most prevalent form of targeted attacks on various IoT devices. Notably, DDoS-ICMP Flood attacks were the most frequent, accounting for 72.75% of all observed attacks. Conversely, Web-based Uploading Attacks were the least common, representing less than 0.01% of the total attacks.

This observation suggests a significant need to effectively enhance security solutions to counteract DDoS threats.

Improvements in defensive strategies will be crucial for mitigating the impact of these attacks on network systems and maintaining cybersecurity integrity.

*Figure 2 Attacks % per Category and Type*

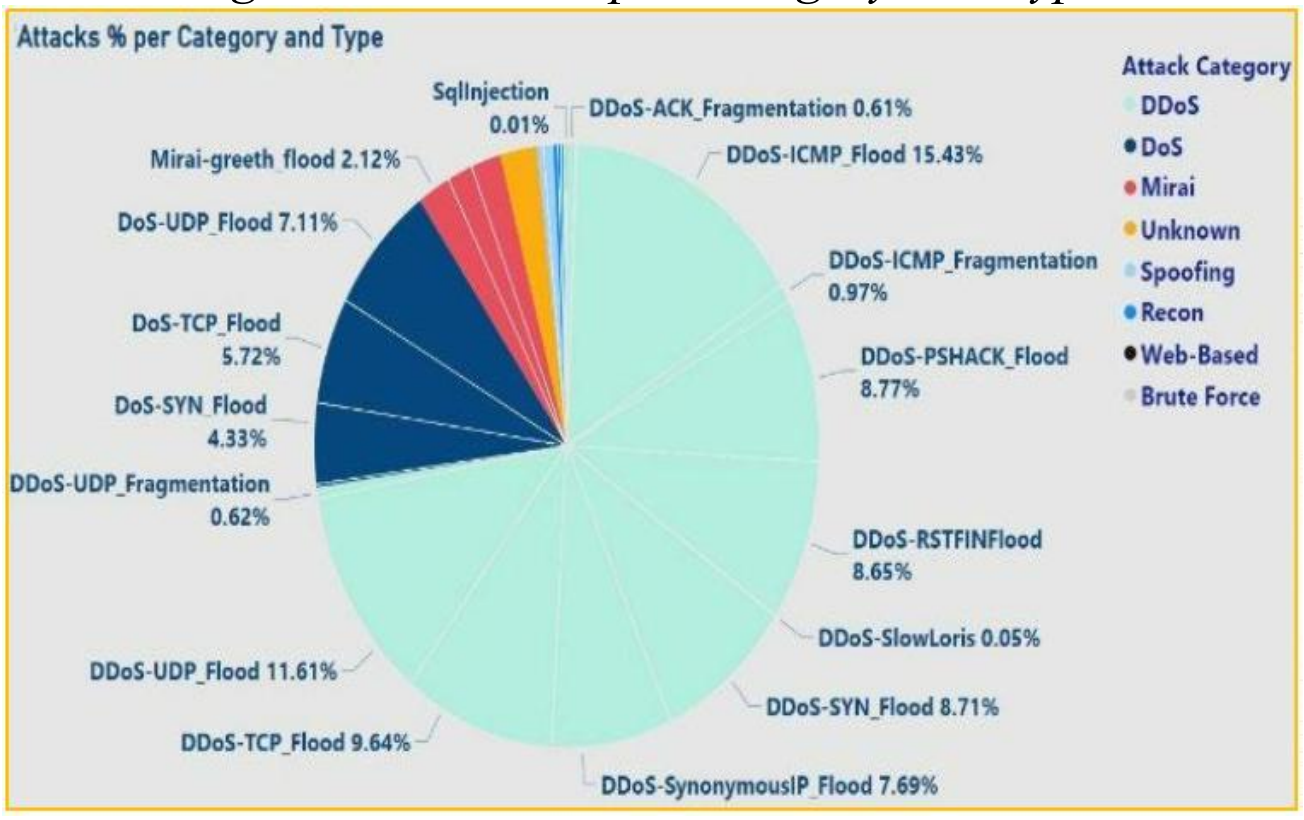

## 6.2 Distribution of Targeted Attacks by Network Protocol

The second visualization, "Figure 3 - Protocol Type % Distribution," was created using Microsoft Excel and presented as a pie chart. This chart illustrates the distribution of targeted attacks across different network protocols within the IoT ecosystem. The parameters utilized include network protocol types and attack counts, enabling the identification of the most and least frequent attack occurrences. The analysis revealed that TCP (Transmission Control Protocol) was the most targeted network protocol, accounting for 47% of all network attacks. This suggests that TCP, as a commonly used but vulnerable protocol, requires extensive security measures and attention to mitigate risks effectively. In contrast, SSH (Secure Shell) and SMTP (Simple Mail Transfer Protocol) protocols were among the least targeted, indicating lower occurrences and potentially higher security within the IoT ecosystem than TCP. Web-based Uploading Attacks were particularly rare, representing less than 0.01% of the total attacks. This observation underscores the necessity of focusing security enhancements on TCP, given its high susceptibility to attacks.

*Figure 3 - Protocol Type % Distribution*

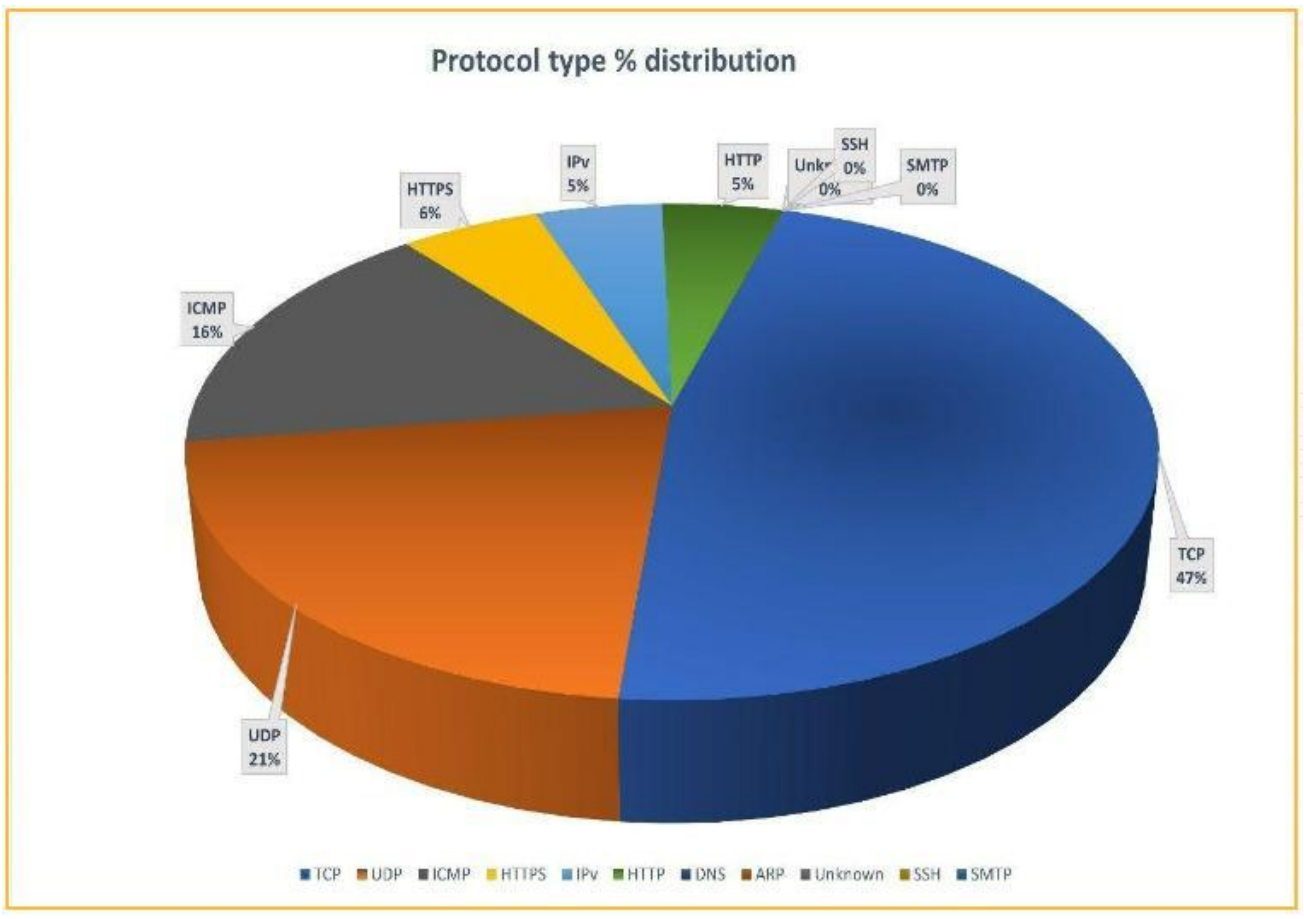

## 6.3 Flag Analysis (Packet Flow) per Attack

The third visualization, "Figure 4 - Flag Usage Graph" is presented as a bar chart created using Microsoft Excel. It provides a detailed view of the frequency and distribution of TCP flag usage across different attack scenarios within the IoT ecosystem. The graph visualizes counts for each flag type—FIN, SYN, PSH, ACK, and ECE[1]- across various attacks, quantitatively assessing flag utilization in network security incidents.

The analysis of the chart highlights significant findings:

- **High Frequency of SYN and ACK Flags**: Particularly noticeable in scenarios labeled as "DDoS-TCP-Flood," where SYN and ACK flags are predominantly used. This usage pattern suggests these flags are critical in facilitating high-volume connection requests, characteristic of flooding attacks aimed at overwhelming network resources.
- **Usage of FIN Flags**: The "Benign Traffic" and "DDoS-TCP-Flood" show substantial use of the FIN flag, which indicates operations involving the initiation or termination of TCP connections. These operations can either simulate normal user behavior or contribute to the exhaustion of server resources.
- **ECE Flag Utilization**: Although less prominent, using the ECE flag in certain attacks highlights its role in congestion notification, which could be exploited in sophisticated network disruption strategies.

*Figure 4 - Flag Usage Graph*

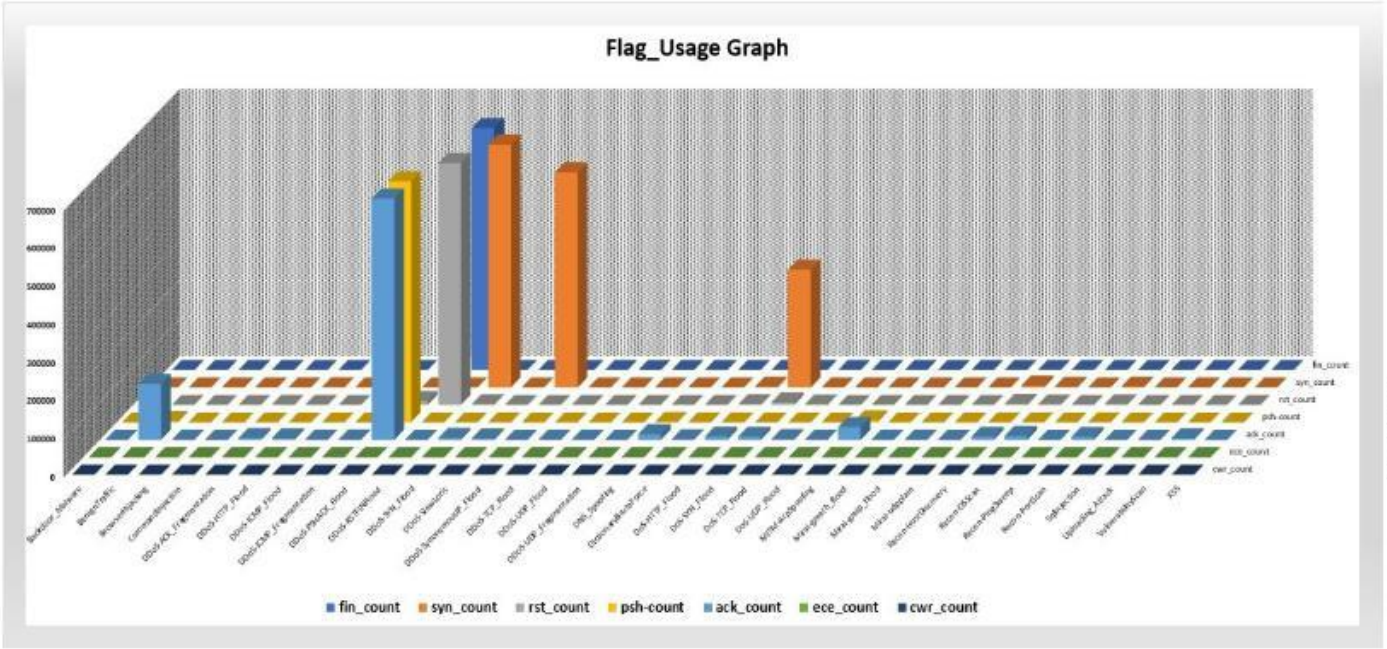

## 6.4 Attack Behavior

The visualization, "Figure 5 - Attack Behavior," was created using Microsoft Excel and presented as a bar chart. This chart illustrates the distribution and frequency of various packet types across different attack categories within the cybersecurity landscape. The parameters utilized include attack categories, total packets, total TCP packets, total UDP packets, SYN packets, FIN packets, and ACK packets. This enables the identification of the most prevalent and significant forms of network behavior during attacks.

- **High Volume of TCP Traffic**: A noticeably higher volume of total packets and total TCP packets associated with DDoS and DoS attacks. This indicates that these attack types are characterized by high levels of network traffic, which is a hallmark of their disruptive capability. The large bars for DDoS and DoS in the 'total_packets' and 'total_tcp_packets' categories underscore the intense use of the TCP protocol to execute these attacks, aiming to overwhelm network resources.

- **Less Targeted Networks:** Attacks categorized as Web-based and Brute Force show significantly lower packet volumes across all types. This suggests that these types of attacks might not rely as heavily on volume to achieve their objectives or are less prevalent compared to the more aggressive DDoS and DoS attacks.
- **Packet Flags in TCP-based Attacks**: SYN, FIN, and ACK packets, which are crucial for managing TCP sessions, are relatively less frequent across all attacks but show notable presence in the context of TCP-based attacks like DDoS and DoS. This reflects the strategic use of these flags in managing and manipulating TCP connections to disrupt services.

This analysis suggests a significant need to enhance security solutions to counteract high-volume DDoS and DoS attacks effectively. Improvements in defensive strategies, such as advanced traffic filtering, rate limiting, and anomaly detection systems, will be crucial for mitigating the impact of these attacks on network systems and maintaining robust cybersecurity integrity. Moreover, the relatively lower frequency of SYN, FIN, and ACK packets in other attack types, like Web-based and Brute Force attacks, indicates different mechanisms of attack, which may require tailored security measures to detect and prevent such threats effectively.

*Figure 5 – Attack Behavior*

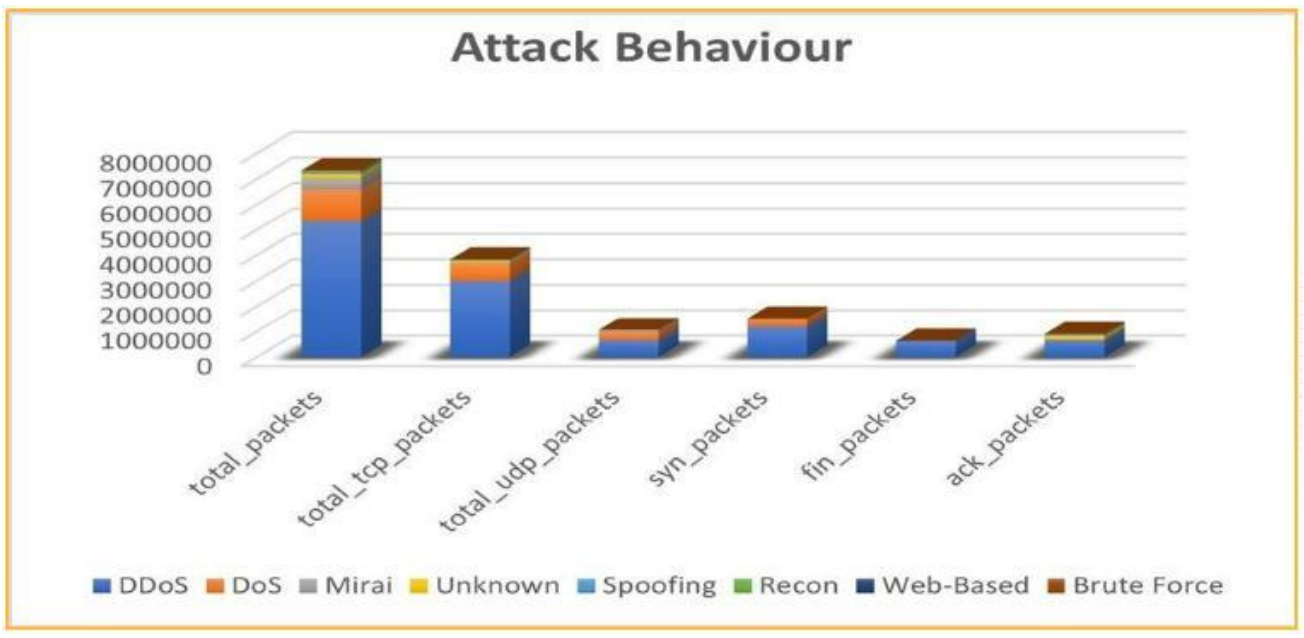

## 6.5 Comparison: HTTP vs. HTTPS Attacks

The visualization "Figure 6 - HTTPS vs HTTP Attacks" is presented as a bar chart created using Tableau. It underscores that while HTTPS is generally more secure due to its encryption protocols, it still encounters a significant volume of traffic, much of which is benign. HTTPS records over 100k in benign traffic incidents, as opposed to less than 10k in HTTP, indicating its widespread use and robust monitoring mechanisms to distinguish and record legitimate network activities. This highlights the importance of considering the volume of benign traffic in security analyses as it reflects the effective utilization of HTTPS for secure communications.

However, the presence of benign traffic does not negate the occurrence of actual attacks. While DDoS Flood attacks are less frequent in HTTPS than HTTP, suggesting the encrypted nature of HTTPS deters or hinders the effectiveness of such attacks, it's crucial to note that HTTPS is not immune to security threats. The analysis still demonstrates a necessity for continuous enhancements in security practices to mitigate potential vulnerabilities in HTTPS and to protect against evolving threats. This includes strengthening the response to less frequent, yet potentially severe, attacks such as SQL Injections and Cross-Site Scripting, which occur at lower rates across both protocols.

*Figure 6 – HTTPs vs HTTP Attacks*

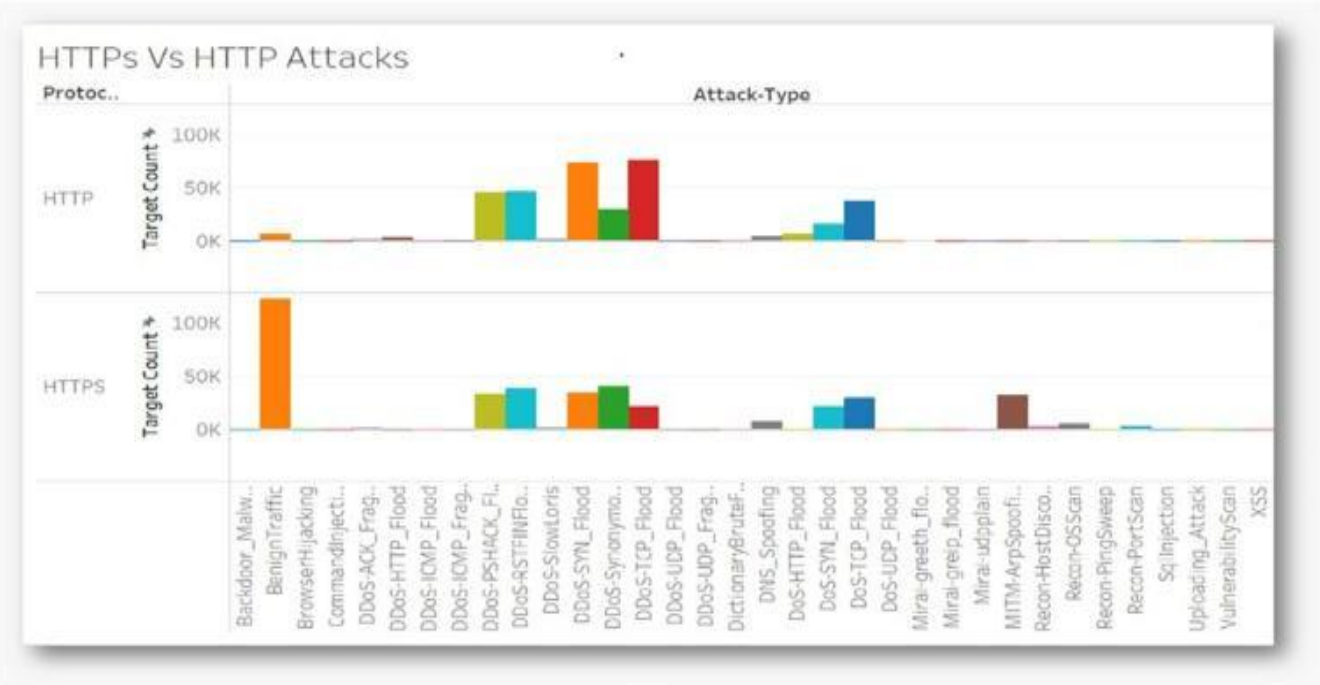

## 7 . Conclusion

In conclusion, our analysis, "Large-Scale Attacks Data Analysis in IoT Environments using Hadoop," addresses critical security challenges in the rapidly evolving Internet of Things (IoT) landscape. By leveraging the extensive and novel CICIoT2023 dataset, our study comprehensively analyzes the myriad threats that besiege IoT devices. Utilizing Apache Hadoop and Hive for big data processing and employing visualization tools like Microsoft Power BI, Tableau, and Excel, we have transformed complex data into actionable insights that accentuate the vulnerabilities and attacks within the IoT ecosystem.

Our findings underscore the continuous need for robust cybersecurity measures, as the standardization and interoperability issues within IoT still pose significant threats to its secure deployment. Our analysis of network protocols and TCP flag usage within various attack scenarios reveals that, despite technological advances, fundamental vulnerabilities like those in TCP can still be exploited through sophisticated cyber-attacks. Moreover, the comparison between HTTP and HTTPS protocols illustrates that while encryption offers a layer of security, no system is impervious to attacks.

1. "FIN, SYN, PSH, ACK, and ECE" are the TCP flags within the TCP header that manage the state and flow of data in a network: Finish, Synchronize, Push, Acknowledge, and Explicit Congestion Notification Echo.